\documentclass[11pt]{article}
\usepackage{amsmath,amsthm,amssymb,graphicx}
\usepackage[margin=1in]{geometry}
\usepackage{amssymb}
\usepackage{amsmath}
\usepackage{subfigure}
\usepackage{amsfonts}
\usepackage{hyperref}
\usepackage{tikz}
\usepackage{booktabs}
\usepackage[all]{xy}

\usepackage{algorithm}
\usepackage{algorithmic}
\theoremstyle{plain}

\title{Persistent Quantum Memory in Iterated Lifts}

\author{Hartosh Singh Bal}
\date{}

\begin{document}
\maketitle

\begin{abstract}

We study quantum coherence in continuous-time quantum walks on perfect graphs generated by the symmetric lift ${\mathrm{HL}}'_2(G)$, a canonical, unweighted, undirected construction defined as the line graph of a bipartite double cover of $G$. This lift acts as both a coherence-preserving and coherence-inducing transformation: it preserves and scales structured quantum interference in highly symmetric base graphs, and induces sustained coherence in random or weakly structured ones. 

In small graphs such as $K_4$, $K_5$, and the Petersen graph, where quantum walks exhibit sharp revivals and high return probability, repeated $\mathrm{HL}'_2$ lifting produces towers of perfect graphs with thousands to tens of thousands of vertices that retain periodic or quasi-periodic coherence. When applied to random regular or Erd\H{o}s--R\'enyi graphs with flat or decaying return behavior, the lift introduces structured interference and significant amplification of mean and peak return probabilities. 

To quantify these effects, we evaluate standard coherence metrics from quantum resource theory, including inverse participation ratio (IPR), purity, relative entropy of coherence, and the logarithmic coherence number. These measures confirm that $\mathrm{HL}'_2$ lifting delocalizes eigenstates, increases coherence entropy, and expands the basis support of quantum states. These results demonstrate that $\mathrm{HL}'_2$ is a scalable and structurally grounded mechanism for organizing quantum interference, and introduce a new family of perfect graphs that support long-time quantum coherence without spectral tuning or engineered weights. 

\end{abstract}

\section{Introduction}

Quantum coherence---the ability of a quantum state to interfere with itself and return to its original configuration---is a delicate phenomenon. In most systems, especially those that are large or structurally complex, coherence rapidly decays due to dispersion, spectral irregularity, or environmental noise. Even in graph-based models of continuous-time quantum walks, coherence is typically short-lived unless the graph is highly regular or carefully engineered for perfect state transfer.

A well-known exception is the complete graph $K_4$, which is small, symmetric, and has a simple integer spectrum. When a quantum walk is initialized at a node of $K_4$, the evolution is perfectly periodic: the state returns to its origin at regular intervals with probability 1. This provides a textbook demonstration of full quantum coherence---but only at the smallest scale.

In this paper, we demonstrate that the same type of coherent behavior can be preserved, and even induced, in far larger graph systems using a canonical graph lift known as the symmetric lift ${\mathrm{HL}}'_2(G)$~\cite{Bal_perfect}. Defined combinatorially as the line graph of the bipartite double cover of $G$, this transformation produces a perfect, claw-free, and unweighted graph whose structure supports harmonic interference and return. By recursively applying this lift, we generate a tower of graphs in which coherence either persists or emerges, even when starting from graphs that are incoherent at the base level.

For highly coherent base graphs such as $K_4$, $K_5$, and the Petersen graph, we show that mathrm{HL}$'_2$ lifts preserve oscillatory return structure under scaling of two or more orders of magnitude in graph size. Even at level 3 (with thousands of vertices), and level 5 (with over 50,000 vertices), we observe persistent, high-amplitude return probability peaks. For weakly coherent or random base graphs such as 3-regular graphs or Erd\H{o}s--R\'enyi models, $\mathrm{HL}'_2$ lifts induce coherence: return profiles exhibit new oscillatory structure, sharp revivals, and significant amplification in mean return.

We simulate continuous-time quantum walks on both base and lifted graphs using sparse Krylov methods, and analyze the resulting dynamics through both visual recurrence plots and structure-sensitive coherence metrics. To our knowledge, this is the first known graph operation that generates scalable, unweighted, and structurally defined graphs that support long-time quantum coherence---not only preserving coherence where it exists, but actively inducing it where it is absent.

\begin{figure}[h]
    \centering
    \includegraphics[width=0.45\textwidth]{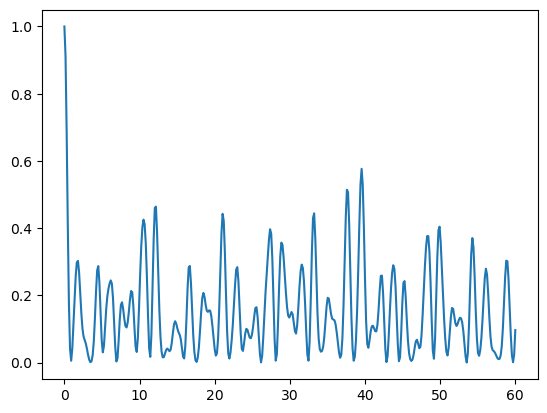}
    \includegraphics[width=0.45\textwidth]{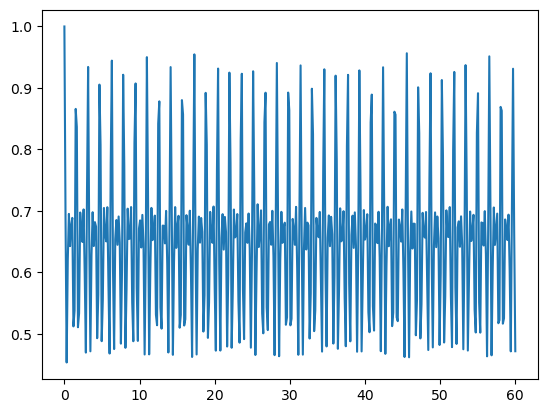}
    \caption{\textbf{Return probability in continuous-time quantum walk on a random regular graph(left) of degree $4$ on $20$ vertices and on its 3-lift ${\mathrm{HL}'_2}^3(G)$ (right). The base graph shows limited structured behavior. The lifted graph, with 4,800 vertices, shows amplified high-amplitude, structured revivals.}}
    \label{fig:rr-lift3}
\end{figure}

To make this phenomenon visually clear, we compare the return probability on a random regular graph of degree four on $20$ vertices with its third symmetric lift ${{\mathrm{HL}}'_2}^3(G)$, which contains $4,800$ vertices. We focus on the third level of the lift tower because it already scales coherence to thousands of nodes while remaining computationally tractable for full quantum walk simulation. The base graph shows limited structure and revival, but as shown in the figure~\ref{fig:rr-lift3}, the lifted graph displays structured, high-amplitude revivals despite its size, acquiring coherence at a scale that is unprecedented for unweighted and undirected graphs.

These findings introduce a new combinatorial route to coherence-aware graph design, with potential applications in analog quantum simulation, passive quantum memory, interference-based processing, and the study of large-scale structured quantum networks.

The symmetric lift ${\mathrm{HL}}'_2(G)$ analyzed in this paper was first introduced as part of a broader theoretical framework for canonical graph lifts in combinational optimization and spectral graph theory~\cite{Bal_perfect}. This framework includes both symmetric and label-sensitive ordered lifts, the latter explored extensively in a companion paper~\cite{Bal_labelcover} focusing on computational complexity and the Unique Games Conjecture. In contrast to those complexity-theoretic investigations, here we demonstrate a striking quantum-information-theoretic capability: the symmetric lift's intrinsic combinational structure supports persistent quantum coherence at unprecedented scales.

To complement our dynamical observations, we also analyze quantum coherence using formal measures from quantum information theory. Specifically, we compute the inverse participation ratio (IPR), purity, relative entropy of coherence, and the logarithmic coherence number, the latter defined via a convex-roof construction based on coherence rank~\cite{xi2018logcoherence}. These metrics, drawn from the resource theory of coherence~\cite{baumgratz2014quantifying}, confirm that symmetric lifts not only preserve return probability structure, but systematically increase spectral delocalization and basis support in a way consistent with structural coherence amplification. This quantitative layer reinforces the visual and dynamical findings and positions mathrm{HL}$'_2$ lifts as scalable, structure-preserving coherence amplifiers within the broader landscape of quantum resource theory.

\section{Background and Definitions}

We study quantum walks on a family of recursively defined graphs derived from the symmetric lift operation \({\mathrm{HL}}'_2(G) \), introduced and analyzed in our previous work on perfect quotient graphs \cite{Bal_perfect}. These graphs are constructed via a combinatorial transformation that preserves and enhances many desirable structural and spectral properties.

\subsection{The Symmetric Lift \({\mathrm{HL}}'_2(G) \)}

Let \( G = (V, E) \) be a finite simple undirected graph. Construct a bipartite graph \( B_{\mathrm{HL}'}(G) \) as follows:
\begin{itemize}
    \item The vertex set is again \( V' \sqcup V'' \), two disjoint copies of \( V \).
    \item For each edge \( \{u,v\} \in E \), include two edges: one from \( u' \in V' \) to \( v'' \in V'' \), and one from \( v' \in V' \) to \( u'' \in V'' \).
\end{itemize}
The Symmetric lift \( \mathrm{HL}'_2(G) \) is the line graph \( L(B_{\mathrm{HL}'}(G)) \) of this bipartite graph.

\subsection{Structural Properties}

The symmetric lift \({\mathrm{HL}}'_2(G) \) inherits and strengthens many classical structural properties:

\begin{itemize}
    \item \textbf{Box-Perfect}: As shown in \cite{Bal_perfect}, the symmetric lift is always box-perfect, meaning the clique vs stable set polytope is integral under box constraints.
    \item \textbf{Claw-Free}: Since the graph is a line graph, it contains no induced \( K_{1,3} \).
    \item \textbf{No Odd Holes or Antiholes}: All induced cycles are of even length, and no complement of an induced odd cycle appears.
    \item \textbf{Perfect Graph}: The lift satisfies the strong perfect graph theorem; its chromatic number equals its clique number on all induced subgraphs.
    \item \textbf{Line Graph of a Bipartite Graph}: This implies all of the above, and plays a central role in spectral and combinatorial behavior.
\end{itemize}

\subsection{Degree, Connectivity, and Spectrum}

Let \( G \) be a connected, \( d \)-regular graph with \( n \) vertices and \( m \) edges. Then:

\begin{itemize}
    \item \( {\mathrm{HL}}'_2(G) \) has \( 2m \) vertices and is \( 2d - 2 \)-regular.
    \item If \( G \) is non-bipartite, \({\mathrm{HL}}'_2(G) \) is connected.
    \item If \( G \) is bipartite, \({\mathrm{HL}}'_2(G) \) has exactly two connected components.
    \item The spectrum of \({\mathrm{HL}}'_2(G) \) is given by:
    \[
    \{2d - 2 - \lambda_i \mid \lambda_i \in \mathrm{Spec}(G)\} \cup \{2d - 2 + \lambda_i \}
    \]
    That is, the eigenvalues of \( G \) get transformed to symmetric intervals around \( d - 2 \), and the minimal eigenvalue is always at least \( -2 \).
\end{itemize}

\subsection{Growth in the $\mathrm{HL}'_2$ Tower}

Starting from the base graph \( G_0 = K_4 \), we define:
\[
G_1 = {\mathrm{HL}}'_2(K_4), \quad G_2 = {\mathrm{HL}}'_2(G_1), \quad \dots, \quad G_n = {\mathrm{HL}}'_2(G_{n-1})
\]
The sequence \( G_n = {{\mathrm{HL}}'_2}^n(K_4) \) forms a deterministic tower of perfect graphs with recursively defined structure.

The following table summarizes the growth of the tower up to level 5:
Let \( V_r \) and \( E_r \) denote the number of vertices and edges of \( G_r \), respectively. Then:
\[
V_{r+1} = 2 E_r = V_r \cdot d_r
\qquad\text{and}\qquad
E_{r+1} = \frac{V_{r+1} \cdot d_{r+1}}{2}
\]

Starting from \( V_0 = 4 \), \( E_0 = 6 \), and \( d_0 = 3 \), the sequence unfolds as:

\begin{center}
\begin{tabular}{|c|c|c|c|}
\hline
Level \( r \) & \( V_r \) & \( d_r \) & \( E_r \) \\
\hline
0 & 4 & 3 & 6 \\
1 & 12 & 4 & 24 \\
2 & 48 & 6 & 144 \\
3 & 288 & 10 & 1,440 \\
4 & 2,880 & 18 & 25,920 \\
5 & 51,840 & 34 & 17,62,560 \\
\hline
\end{tabular}
\end{center}

This exponential growth arises from the doubling of edges at each lift, and the degree rule \( d \mapsto 2d - 2 \). Despite this growth, all $\mathrm{HL}'$ lifts remain perfect, sparse, and algorithmically tractable.

\begin{table}[h]
\centering
\begin{tabular}{c|c}
Level \( n \) & Distinct Eigenvalues of \( \mathrm{{HL}}'^n_2(K_4) \) \\
\hline
0 & \( \{ 3, -1 \} \) \\
1 & \( \{ 4, 2, 0, -2 \} \) \\
2 & \( \{ 6, 4, 2, 0, -2 \} \) \\
3 & \( \{ 10, 8, 6, 4, 2, 0, -2 \} \) \\
4 & \( \{ 18, 16, 14, 12, 10, 8, 6, 4, 2, 0, -2 \} \) \\
5 & \( \{ 34, 32, 30, 28, 26, 24, 22, 20, 18, 16, 14, 12, 10, 8, 6, 4, 2, 0, -2 \} \)
\end{tabular}
\caption{Distinct eigenvalues (spectral spread) of the lifted graphs \( {\mathrm{HL}'_2}^n(K_4) \) up to level 5. Each spectrum contains the previous level's spectrum.}
\end{table}

\subsection{Return Probability as a Measure of Coherence}

To study quantum coherence on graphs, we use the framework of continuous-time quantum walks~\cite{childs}. Given a graph \( G \) with adjacency matrix \( A \), the quantum state \( \psi(t) \in \mathbb{C}^{|V|} \) evolves according to Schrödinger’s equation:
\[
\psi(t) = e^{-itA} \psi(0),
\]
where \( \psi(0) \) is the initial state and \( t \) is time. We take \( \psi(0) \) to be localized at a single vertex \( v \in V(G) \), i.e., \( \psi(0) = \delta_v \), the standard basis vector.

A key quantity of interest is the \emph{return probability}:
\[
P_{\mathrm{return}}(t) = |\langle \psi(0), \psi(t) \rangle|^2 = |\psi_v(t)|^2,
\]
which measures the likelihood that the quantum walker returns to its starting point at time \( t \). This quantity is bounded between 0 and 1 and captures the amount of constructive interference at the origin.

We interpret sustained or recurring peaks in \( P_{\mathrm{return}}(t) \) as a sign of coherence. A system that regularly revives to its initial state reflects spectral structure and global phase alignment. Conversely, in most large or irregular graphs, the return probability rapidly decays and remains near its long-time average, typically close to \( 1/|V| \). This flattening signals dephasing and loss of coherence.

The $\mathrm{HL}'_2$ tower presents an intriguing case: although each lift greatly increases the number of vertices and the degree, the return probability often remains high and oscillatory over long timescales. We use this behavior as a proxy for coherence and investigate how it evolves with depth in the tower.

\subsection{Simulation Framework}

To evaluate quantum coherence on large graphs from the $\mathrm{HL}'_2$ tower, we simulate continuous-time quantum walks using a sparse numerical method. Direct matrix exponentiation is infeasible for large graphs, so we use the Krylov subspace approximation to compute the action of the matrix exponential on a localized initial state.

\paragraph{Numerical Method.} Given the adjacency matrix \( A \) of a graph \( G \) and an initial state vector \( \psi(0) \), we compute
\[
\psi(t) = e^{-itA} \psi(0)
\]
at discrete time steps \( t \in [0, T] \) using the \texttt{expm\_multiply} method from \texttt{SciPy}. This approach constructs a low-dimensional Krylov basis around \( \psi(0) \) and approximates the exponential in that basis, avoiding the need to store or manipulate the full matrix exponential.

\paragraph{Graph Construction.} All simulations begin with a base graph \( G_0 \). Each lift is generated recursively using the symmetric $\mathrm{HL}'_2$ operator:
\[
G_{n+1} = \mathrm{{HL}}'_2(G_n),
\]
which we implement via edge-based construction over ordered pairs.

\paragraph{Subgraph Sampling.} For levels beyond \( n = 3 \), the size of \( G_n \) can become large. To keep simulations tractable while capturing local coherence behavior, in some cases we extract connected subgraphs of fixed size using a randomized breadth-first search. Starting from a uniformly sampled node, we grow a connected region containing up to \( N_{\max} \) nodes, where \( N_{\max} \in \{3000\} \) depending on the experiment.

\paragraph{Simulation Parameters.} For each subgraph, we simulate the return probability from the chosen starting node over a time interval \( [0, T] \), with \( T \) typically ranging from 30 to 120. We discretize the interval into 400–800 steps and record the return amplitude \( P_{\mathrm{return}}(t) \) at each point.

All simulations were conducted using standard \texttt{NumPy} and \texttt{SciPy} routines, with graph construction in \texttt{NetworkX} and visualization in \texttt{Matplotlib}. The Python pipeline is efficient enough to simulate graphs with up to tens of thousands of vertices in seconds to minutes.

\subsection{Complete Graphs: $K_4$}

\begin{figure}[H]
    \centering
    \subfigure[$K_4$]{\includegraphics[width=0.3\textwidth]{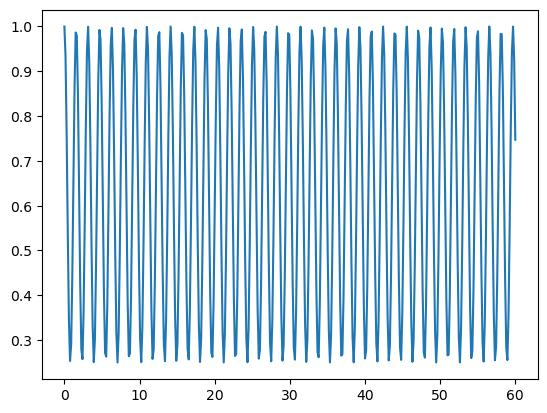}}
    \subfigure[${\mathrm{HL}'_2}(K_4)$]{\includegraphics[width=0.3\textwidth]{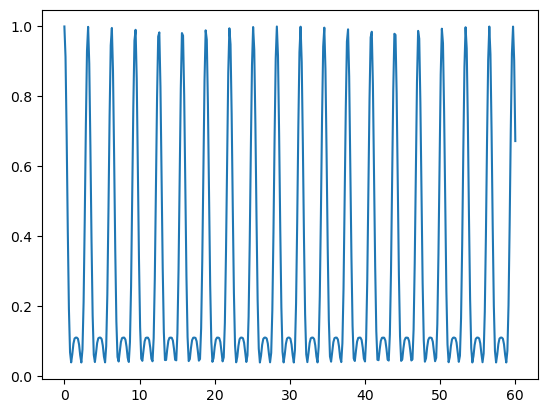}}
    \subfigure[${\mathrm{HL}'_2}^2(K_4)$]{\includegraphics[width=0.3\textwidth]{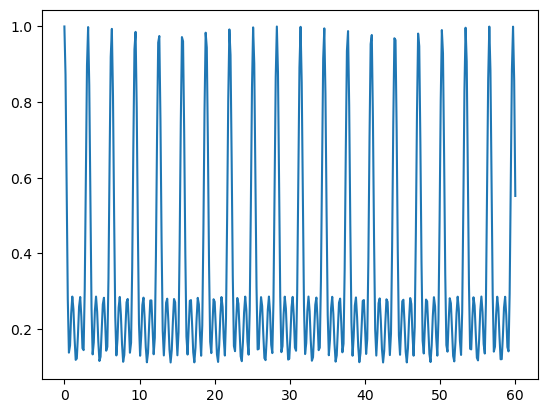}}\\[1ex]
    \subfigure[${\mathrm{HL}'_2}^3(K_4)$]{\includegraphics[width=0.3\textwidth]{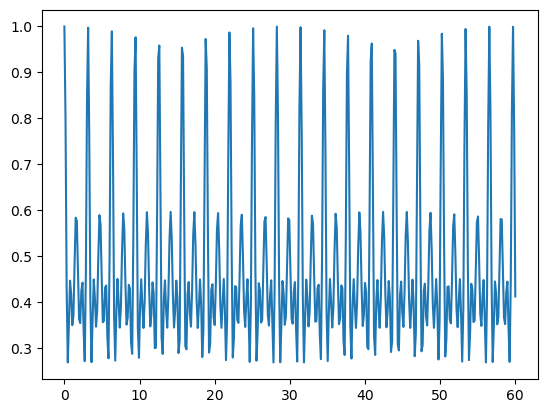}}
    \subfigure[${\mathrm{HL}'_2}^4(K_4)$]{\includegraphics[width=0.3\textwidth]{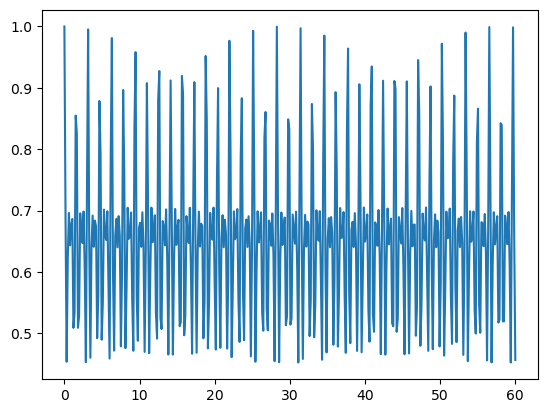}}
    \caption{\textbf{Return probability in continuous-time quantum walk on $K_4$ and its successive symmetric lifts. Coherence persists across all levels, including ${\mathrm{HL}'_2}^4(K_4)$ with 2,880 vertices and degree 18.}}
    \label{fig:k4-lift-grid}
\end{figure}

We begin with the complete graphs $K_4$ and $K_5$, which are known to support perfect quantum coherence at small scale. As shown in Figure~\ref{fig:k4-lift-grid}, $K_4$ exhibits exact periodic return with probability 1, while its fourth symmetric lift $\mathrm{{HL}}'^4_2(K_4)$---with 2,880 vertices---preserves high-amplitude, regular revivals over long time scales.

We also observe the same qualitative behavior in $K_5$ and its symmetric lifts: periodic return is preserved under recursive lifting, and the coherence profile closely resembles that of $K_4$, scaled in size and period. For clarity, we focus in this paper on four representative examples—$K_4$, the Petersen graph, a random 3-regular graph, and a random Erdős–Rényi graph—to illustrate both preservation and emergence of quantum coherence.

\begin{figure}[H]
    \centering
    \includegraphics[width=0.45\textwidth]{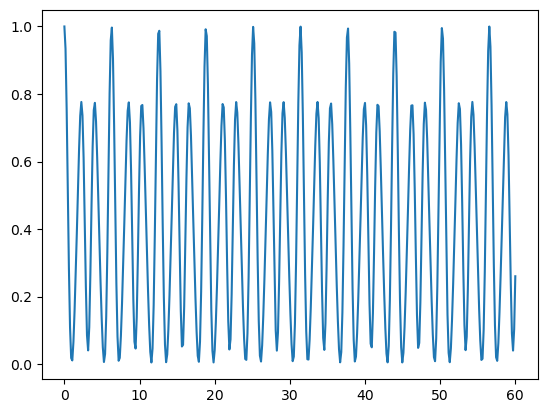}
    \includegraphics[width=0.45\textwidth]{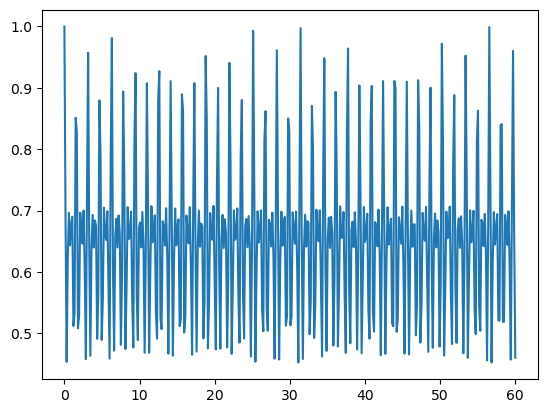}
    \caption{\textbf{Return probability in continuous-time quantum walk on the Petersen graph (left) and its fourth symmetric lift ${{mathrm{HL}}'_2}^4(\text{Petersen})$ (right). Coherence remains visible and periodic at scale, even with over 7,000 vertices in the lifted graph.}}
    \label{fig:petersen-lift4}
\end{figure}

\begin{figure}[H]
    \centering
    \includegraphics[width=0.45\textwidth]{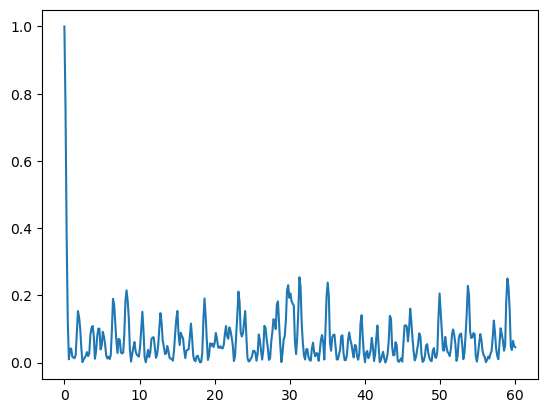}
    \includegraphics[width=0.45\textwidth]{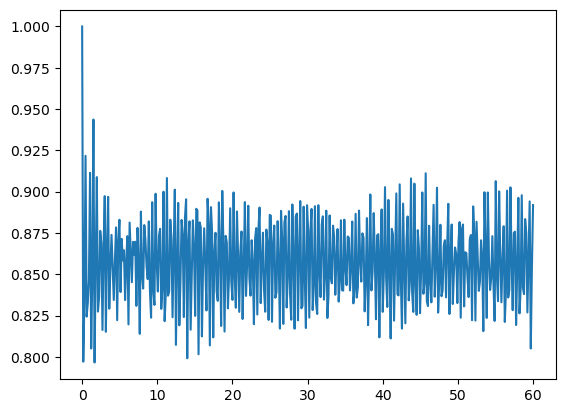}
    \caption{\textbf{Return probability in continuous-time quantum walk on a random Erdős–Rényi graph $G(40, 0.3)$ (left), and its second symmetric lift ${mathrm{HL}'_2}^2$ (right). The base graph exhibits no periodicity, while the lifted graph shows pronounced coherence and high-amplitude return.}}
    \label{fig:gnp-coherence}
\end{figure}

\begin{figure}[H]
    \centering
    \begin{minipage}[b]{0.48\textwidth}
        \centering
        \includegraphics[width=\textwidth]{rr4_20_base.png}
        \caption{Base (20 nodes)}
    \end{minipage}%
    \hfill
    \begin{minipage}[b]{0.48\textwidth}
        \centering
        \includegraphics[width=\textwidth]{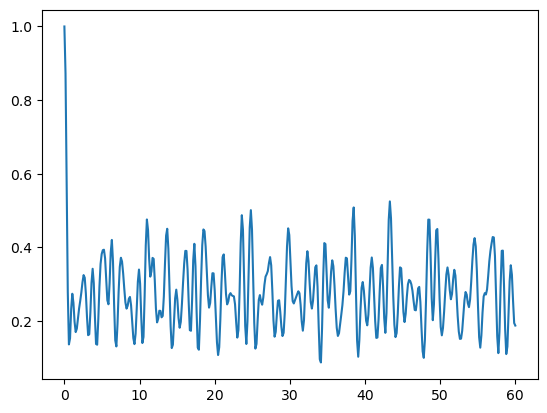}
        \caption{$\mathrm{HL}'_2$ Lift (80 nodes)}
    \end{minipage}
    
    \vspace{1ex}
    
    \begin{minipage}[b]{0.48\textwidth}
        \centering
        \includegraphics[width=\textwidth]{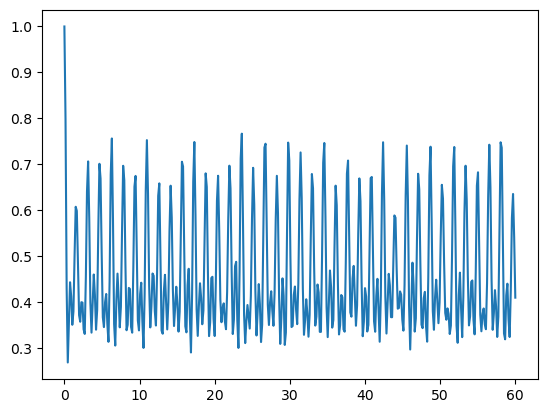}
        \caption{${\mathrm{HL}'_2}^2$ Lift (480 nodes)}
    \end{minipage}%
    \hfill
    \begin{minipage}[b]{0.48\textwidth}
        \centering
        \includegraphics[width=\textwidth]{rr4_20_lift3.png}
        \caption{${\mathrm{HL}'_2}^3$ Lift (4800 nodes)}
    \end{minipage}
    
    \caption{\textbf{Return probability in continuous-time quantum walk on a random 4-regular graph and its symmetric lifts. Coherence is absent in the base but rapidly emerges across lifts, with ${\mathrm{HL}'_2}^3$ showing structured, high-contrast revivals.}}
    \label{fig:rr4-grid}
\end{figure}

\section{Structural Indicators of Coherence}

To explain the preservation and emergence of coherence across symmetric lifts, we examine how key structural properties of the lifted graphs evolve. Table~\ref{tab:coherence-metrics} summarizes this evolution for an Erdős–Rényi base graph $G(20, 0.1)$ and its successive symmetric lifts ${\mathrm{HL}}'_2$, ${{\mathrm{HL}}'_2}^2$, and ${{\mathrm{HL}}'_2}^3$.

\begin{table}[h]
\centering
\small
\begin{tabular}{|c|r|r|r|r|r|r|r|}
\hline
\textbf{Lift Level} & \textbf{Vertices} & \textbf{Edges} & \textbf{Tr$(A^4)$} & \textbf{Tr$(A^4)/n$} & \textbf{Avg Clust.} & \textbf{Triangles} & \textbf{Triangles/Vertex} \\
\hline
Base       & 20   & 21    & 222      & 11.10   & 0.1683 & 3     & 0.15 \\
$\mathrm{HL}'_2$    & 42   & 86    & 1820     & 43.33   & 0.6236 & 52    & 1.24 \\
${\mathrm{HL}'_2}^2$  & 172  & 624   & 29696    & 172.65  & 0.4522 & 660   & 3.84 \\
${\mathrm{HL}'_2}^3$  & 1248 & 8592  & 1152128  & 923.18  & 0.4721 & 18376 & 14.72 \\
\hline
\end{tabular}
\caption{\textbf{Structural coherence metrics for $G(20, 0.1)$ and successive $\mathrm{HL}'_2$ lifts. Trace $(A^4)$ counts closed 4-step walks; clustering and triangle counts reflect local path density. Triangle density per vertex highlights the amplification of walk overlap.}}
\label{tab:coherence-metrics}
\end{table}

These results highlight how symmetric lifting transforms the walk space of a graph. Even starting from a sparse and incoherent base, $\mathrm{HL}'_2$ lifts rapidly enrich the structure with short closed walks, overlapping paths, and dense local neighborhoods—all of which support constructive interference in quantum dynamics. The trace of $A^4$ grows by several orders of magnitude across the tower, indicating a sharp increase in the number of short coherent walk cycles that contribute significantly to early-time return probability. The normalized trace $\mathrm{Tr}(A^4)/n$ shows that this growth is not merely a consequence of size, but reflects an increase in local coherence structure per vertex. 

The average clustering coefficient increases and stabilizes at a high level, signaling the formation of tightly knit neighborhoods where paths repeatedly overlap~\cite{godsil}. Meanwhile, the triangle count and especially triangle density rise dramatically, from fewer than one per node to over fourteen per node in ${{\mathrm{HL}}'_2}^3(G)$. This dense mesh of short loops and overlapping paths provides a structural foundation for the empirical coherence observed in lifted graphs: walk dynamics become geometrically reinforced, enabling return amplitudes to persist and sharpen across the $\mathrm{HL}'_2$ tower.

\section{Coherence Metrics and Structural Amplification}

Recent work by Scholes~\cite{scholes2023coherence} has shown that large-scale quantum coherence can persist in disordered $k$-regular graphs. Using standard resource-theoretic metrics—\textit{purity}, \textit{inverse participation ratio} (IPR), and \textit{relative entropy of coherence}—that quantify delocalization and deviation from classical mixedness, Scholes demonstrates that coherence may survive even under random edge deletion. His results emphasize the \textit{robustness} of coherence in statistical ensembles of graphs.

In contrast, our focus is on a deterministic and recursive lifting process: the symmetric $\mathrm{HL}'_2$ operation. Rather than relying on spectral averages over random graphs, we explicitly construct lifted graphs in which coherence is not only preserved but \textit{amplified}. To validate this, we apply the same coherence metrics used by Scholes to ${{\mathrm{HL}}'_2}^r(G)$ at levels $r = 0$ to $3$, for representative base graphs $G$.

These are standard quantum information measures of coherence, as introduced in Plenio et al.~\cite{baumgratz2014quantifying}, including the relative entropy of coherence, inverse participation ratio (IPR), and purity.

For each level, we compute:
\begin{itemize}
    \item The average IPR across the top 5 eigenstates of the adjacency matrix;
    \item The purity of the corresponding mixed-state density matrix;
    \item The relative entropy of coherence, measuring deviation from the decohered (diagonal) state;
    \item The mean, peak, and standard deviation of the return probability under continuous-time quantum walk evolution.
\end{itemize}

For each graph and lift level, we select the top five eigenvectors of the adjacency matrix (by eigenvalue magnitude) to form a mixed-state density matrix:
\[
\rho = \frac{1}{5} \sum_{i=1}^5 |\psi_i\rangle\langle\psi_i|,
\]
where each $\psi_i$ is an eigenvector normalized in the standard (computational) basis. The purity and relative entropy of coherence are computed with respect to this basis. The decohered reference state $\rho_{\text{diag}}$ is obtained by zeroing all off-diagonal entries of $\rho$ in this same basis.

All coherence metrics—IPR, purity and relative entropy, —are computed using the top $k=5$ eigenstates of the adjacency matrix for each graph, unless stated otherwise. For small graphs like $K_4$, we verify that using the full eigenbasis yields a purity of exactly $0.25$, consistent with a uniform mixture over four orthonormal eigenstates. The relative entropy remains close to zero in this case, as all eigenvectors are nearly diagonal in the computational basis. This confirms that our top-$k$ truncation is reasonable for larger graphs and captures the most coherent part of the spectrum, though full-spectrum results for small graphs provide a more exact baseline.

We report results for two graphs:
\begin{enumerate}
    \item The complete graph $K_4$, which is fully symmetric and supports coherent evolution from the outset;
    \item A random 3-regular graph on 20 vertices, which is structurally unstructured and exhibits weak base-level coherence.
\end{enumerate}

\begin{table}[h]
\centering
\caption{\textbf{Coherence Metrics for $K_4$}}
\label{tab:k4_coherence}
\begin{tabular}{cccccccc}
\toprule
Lift & Nodes & IPR & Purity & Rel. Entropy & Mean Return & Peak & Std Dev \\
\midrule
0 & 4     & 0.458 & 0.16 & 0.000 & 0.633 & 1.0 & 0.265 \\
1 & 12    & 0.133 & 0.20 & 1.256 & 0.332 & 1.0 & 0.334 \\
2 & 48    & 0.045 & 0.20 & 3.151 & 0.361 & 1.0 & 0.278 \\
3 & 288   & 0.008 & 0.20 & 5.744 & 0.484 & 1.0 & 0.194 \\
\bottomrule
\end{tabular}
\end{table}

\begin{table}[h]
\centering
\caption{\textbf{Coherence Metrics for 3-Regular Graph (Corrected)}}
\label{tab:r3_coherence}
\begin{tabular}{cccccccc}
\toprule
Lift & Nodes & IPR & Purity & Rel. Entropy & Mean Return & Peak & Std Dev \\
\midrule
0 & 20    & 0.077 & 0.20 & 1.970 & 0.148 & 1.0 & 0.190 \\
1 & 60    & 0.031 & 0.20 & 3.539 & 0.179 & 1.0 & 0.179 \\
2 & 240   & 0.008 & 0.20 & 5.539 & 0.300 & 1.0 & 0.151 \\
3 & 1440  & 0.001 & 0.20 & 8.123 & 0.465 & 1.0 & 0.142 \\
\bottomrule
\end{tabular}
\end{table}

These results highlight two complementary phenomena. For $K_4$, $\mathrm{HL}'_2$ preserves and distributes coherence more uniformly: the IPR falls while relative entropy grows, and the return probability remains high. For the $3$-regular graph, which begins with weak coherence, $\mathrm{HL}'_2$ lifts induce a marked transition: entropy increases from $1.97$ to $8.12$, and return probability rises from $0.15$ to $0.46$.

In the case of $K_4$, the decreasing IPR and increasing entropy reflect that coherence becomes more uniformly distributed across the graph. Despite the increase in size under lifting, return probability remains strong, and the low standard deviation suggests stable revivals.

For the 3-regular graph, which begins with low coherence, we observe a marked increase in both relative entropy and return probability. This indicates that the $\mathrm{HL}'_2$ lifting process not only preserves coherence where it exists, but can induce it in initially incoherent systems, creating structured interference pathways absent in the base graph.

Together, these tables demonstrate that $\mathrm{HL}'_2$ lifting is not merely coherence-preserving but \textit{coherence-amplifying}. Unlike Scholes' random ensemble constructions, our approach provides a scalable and structural mechanism for generating coherent quantum systems—grounded in graph lifting.

We conjecture that the relative entropy of coherence for ${{\mathrm{HL}}'_2}^r(G)$ scales at least logarithmically in the number of vertices, due to the increased complexity and density of interference-supporting cycles in the lifted graph. This suggests that structural lifting may provide a scalable mechanism for engineering coherence in large networks.

\paragraph{Logarithmic Coherence Number.}
In addition to purity and entropy-based metrics, we consider the \textit{logarithmic coherence number} (LC), introduced by Xi and Yuwen~\cite{xi2018logcoherence}. This measure quantifies the minimal number of incoherent basis states required to express a quantum state under convex roof construction. Formally, for a pure state $|\psi\rangle$, LC is defined as:
\[
LC(|\psi\rangle) = \log_2 \text{rank}_{\text{coh}}(|\psi\rangle),
\]
where the coherence rank is the number of non-zero coefficients in the computational basis. For a mixed state $\rho$, LC is defined via convex roof:
\[
LC(\rho) = \min_{\{p_i, |\psi_i\rangle\}} \sum_i p_i \log_2 \text{rank}_{\text{coh}}(|\psi_i\rangle).
\]
Although this is difficult to compute exactly, we approximate it by averaging the log of support sizes (above a threshold) for the top 5 eigenstates of the adjacency matrix of each lifted graph.

It is approximated as:
\[
LC(\rho) \approx \frac{1}{k} \sum_{i=1}^k \log_2 \left| \text{supp}(\psi_i) \right|,
\quad \text{where } \text{supp}(\psi_i) = \{ j : |\psi_i(j)| > 10^{-6} \}.
\]
This measure reflects the average number of basis states required to express dominant eigenstates with nontrivial amplitude and tracks well with relative entropy and IPR in both structured and unstructured graphs.

\vspace{1em}
\begin{table}[h]
\centering
\caption{\textbf{Approximate Logarithmic Coherence Number for $K_4$ and its ${{\mathrm{HL}}'_2}^r(G)$ lifts.}}
\label{tab:k4_log_coherence}
\begin{tabular}{cccccc}
\toprule
Lift & Nodes & Log Coherence & Avg IPR & Rel. Entropy & Mean Return \\
\midrule
0 & 4     & 1.79  & 0.458 & 0.000 & 0.630 \\
1 & 12    & 3.53  & 0.133 & 1.26  & 0.307 \\
2 & 48    & 5.56  & 0.045 & 3.15  & 0.344 \\
3 & 288   & 8.14  & 0.008 & 5.74  & 0.479 \\
\bottomrule
\end{tabular}
\end{table}
\vspace{0.5em}

\begin{table}[h]
\centering
\caption{\textbf{Approximate Logarithmic Coherence Number for a Random 3-Regular Graph and its ${{\mathrm{HL}}'_2}^r(G)$  Lifts.}}
\label{tab:r3_log_coherence}
\begin{tabular}{ccccc}
\toprule
Lift & Nodes & Log Coherence & Avg IPR & Rel. Entropy \\
\midrule
0 & 20    & 4.32  & 0.077 & 1.97 \\
1 & 60    & 5.72  & 0.031 & 3.54 \\
2 & 240   & 7.75  & 0.0079 & 5.54 \\
3 & 1440  & 10.36 & 0.0013 & 8.12 \\
\bottomrule
\end{tabular}
\end{table}

These values exhibit a nearly linear growth in log coherence with lift level, supporting our conjecture that structural lifting amplifies the number of basis components required to represent coherent states.

This trend confirms that $\mathrm{HL}'_2$ lifting not only amplifies dynamical coherence (as seen in return probabilities), but also increases the minimum number of basis states required to represent quantum states — consistent with the structural complexity induced by recursive lifts.

\section{Conclusion}

The results presented in this paper point to a remarkably robust phenomenon: quantum coherence, as measured by return probability and revival structure in continuous-time quantum walks, is not merely preserved but often amplified through successive applications of the symmetric graph lift $\mathrm{{HL}}'_2$. 

Beginning with a variety of base graphs—including highly structured graphs such as $K_4$ and the Petersen graph, as well as unstructured or incoherent ones such as random regular and Erdős–Rényi graphs—we observed that coherence becomes increasingly prominent through recursive lifting. Even graphs with flat or noisy return profiles at the base level developed sharp, periodic revivals by the third or fourth lift. In particular, the appearance of high envelope contrast and persistent revival counts across large ${\mathrm{HL}'_2}^r$ towers (some exceeding 7000 vertices) indicates a kind of emergent coherence intrinsic to the lifting process.

These findings suggest a broader dynamical principle: for a wide class of graphs, there exists an integer $r$ such that ${{\mathrm{HL}}'_2}^r(G)$ exhibits structured quantum coherence, regardless of the coherence behavior of the original graph $G$. This raises the possibility of using symmetric lifts as a general-purpose method for inducing coherence in arbitrary quantum systems modeled by graphs.

The apparent universality and scalability of this behavior merit further investigation. In future work, we hope to formally characterize the conditions under which coherence emerges in ${{\mathrm{HL}}'_2}^r(G)$, quantify the number of lifts required in terms of base graph parameters, and explore potential applications in quantum memory, decoherence suppression, and quantum-inspired algorithms on lifted structures.

The results presented here highlight the remarkable combinational versatility and broader utility of the canonical graph lifting constructions introduced in~\cite{Bal_perfect}. While these symmetric lifts sustain quantum coherence as demonstrated in this work, the complementary ordered lifts have concurrently provided deep insights into computational complexity, particularly concerning Label-Cover approximations and the Unique Games Conjecture~\cite{Bal_labelcover}. Together, these results illustrate that canonical graph lifts provide a robust theoretical platform for applications ranging from quantum information processing to theoretical computer science and combinational optimization.

These lifted graphs can be viewed as explicit answers to the open question posed by Scholes~\cite{scholes2023coherence}, who asks whether one can systematically construct networks that host emergent coherent states in large quantum systems. Our symmetric lift ${\mathrm{HL}}'_2(G)$ provides such a construction: it supports the emergence of delocalized, high-coherence states with predictable structural origins, scalable without tuning, and validated by multiple coherence metrics.

Our construction contrasts with the extensive literature on \emph{perfect state transfer} (PST) in graphs, notably developed by Godsil and collaborators~\cite{godsil}. PST seeks to transmit quantum states between vertices with unit fidelity at specific times, often requiring strict spectral conditions (e.g., rational eigenvalues, paired symmetry) and carefully engineered edge weights or graph products. While PST provides high-fidelity quantum routing in small systems, it is fragile and difficult to scale.

In contrast, the symmetric lift ${\mathrm{HL}}'_2(G)$ achieves persistent coherence across thousands of vertices in a purely unweighted and combinatorially defined manner. Rather than aiming for perfect transfer between nodes, our focus is on maintaining global coherence—as captured by return probability revivals, inverse participation ratio, and entropy metrics—under recursive graph expansion. These lifts induce structured interference even from incoherent base graphs, supporting a form of scalable coherence that complements the exact, localized dynamics of PST.

Future work may investigate whether $\mathrm{HL}'_2$ lifts can support approximate PST, or whether hybrid constructions—such as weighted variants of the lift—could bridge the gap between PST and large-scale coherence-preserving architectures.

\section{Appendix: Code for Lift Construction}
\label{appendix:code}

The Python code used to construct symmetric graph lifts introduced in this paper, as well as to compute the quantum coherence metrics and return probabilities reported in Section~3.5, is available on GitHub at:
\begin{center}
\url{https://shorturl.at/Kg4T1}
\end{center}

\begin{center}
\url{https://shorturl.at/LUsq5}
\end{center}

The repository includes:
\begin{itemize}
    \item Code to generate the $\mathrm{HL}'_2$ lift for any NetworkX graph.
    \item Coherence metric computation using standard quantum information measures:
    \begin{itemize}
        \item Inverse Participation Ratio (IPR),
        \item Purity of the density matrix,
        \item Relative entropy of coherence.
    \end{itemize}
    \item Simulation of return probability using continuous-time quantum walks.
    \item Scripts to generate the tables reported in Section~3.5 for $K_4$ and a random 3-regular graph.
    \item CSV and LaTeX outputs for direct inclusion in this paper.
\end{itemize}

The code is modular and can be adapted to study other graph families, coherence measures, or lifting operations. All experiments in the paper were generated using these scripts with minimal modification.

\vspace{2em}
\noindent\textbf{Author address:} \\
Hartosh Singh Bal \\
The Caravan, Jhandewalan Extn., New Delhi 110001, India \\
\texttt{hartoshbal@gmail.com}

\end{document}